# Predicting CEO Compensation in Non-Controlled Public Corporations with the Canonical Regression Quantile Method

Joseph Haimberg[1] and Stephen Portnoy[2]

December 11, 2020

**ABSTRACT**

*Empirical research on the relationship between CEO compensation and corporate performance has, as of yet, failed to yield meaningful results. Evidence instead points to the decoupling of CEO total compensation (salary, bonus, other payments, and stock options) from corporate performance. Recent work by Barton et al (2017) suggested a predictive model where public corporations engaging in long-term operational and financial goals achieve superior performance and market results in the long-term. The suggested predictive power of this long-termism could provide a quantitative measure of future corporate performance that are tied to current CEO action. If true, this model can be used as a proxy to establish future CEO compensation. Corporate ownership structure can influence CEO compensation to reflect ownership powers rather than corporate performance. Therefore, we limited our research to non-controlled public corporations – those with a single class of shares and without blockholders. The analysis here uses Canonical Regression Quantiles to produce an Index based on past data in order to predict CEO compensation two years ahead (see Portnoy and Haimberg, 2020). The regressions on Index are significant, but not overwhelmingly so. They provide a useful tool to separate over-compensated from under-compensated CEOs, and are a method to determine current CEO compensation.*

***Results:*** *The use of the Canonical Regression Quantiles' Index proved that non-controlled companies that invest in long-termism post superior future performance. The Index indicates that current CEO compensation influences future performance. The Index provides a method for determining CEO pay for the next 1-2 year, and is a useful method to distinguish over/under paid CEOs as an unbiased alternative to the peer groups comparison used by most compensation consultants. This determination is statistically weak, but future research using the Canonical Regression Quantiles with a larger data set may lead to increased sensitivity and a powerful unbiased method for replacing compensation consultants who are responsible to the decoupling of CEO compensation and corporate performance.*

## 1. Introduction

In 2002, CEO Dennis Kozlowski and his CFO were accused of stealing approximately $600 million from Tyco (NYSE: TYC). Kozlowski gained notoriety for his extravagant spending of this money on personal expenses "including yachts, fine art, estate jewelry, luxury apartments and vacation estates, personal business ventures and investments, all unrelated to Tyco" (Securities and Exchange Commission 2002). Jurors saw a 30-minute video of the $2 million 40[th] birthday party for Kozlowski's wife in Sardinia—all paid for by Tyco—and sentenced Kozlowski to serve a minimum of eight years and four months, with a maximum of 25 years (CNN 2003). Kozlowski's extreme case garnered outrage and refocused attention on why CEOs are making so much money, legally or not, and why pay and corporate performance do not seem to correlate.

---

[1]Department of Public Affairs & Policy
Mark O. Hatfield School of Government
Portland State University
Correspondence email: yossi@pdx.edu

[2]Professor, Department of Statistics
University of Illinois at Urbana-Champaign

Countless research has examined the CEO compensation issue through the Principal-Agent theory, which implies a relationship where one party, or "agent," acts for another party, the "principal." This interdependent relationship is widespread when asymmetric information exists; other examples include doctor-patient, lawyer-client, advisor-client, teacher-student and auto mechanic-client relationships. (Mitnick 1975; Jensen and Meckling 1976). The parties are free to negotiate the employment relationship terms and conditions at-will, and the amount paid upon completion—or, the agency's cost. Theoretically, the same agency approach exists between shareholders and the CEO, as shareholders are the principals and the CEO acts as their agent, performing specified tasks on behalf of the principals who pay the agency costs in the form of a pay package (Jensen and Meckling 1976). Given the broadly dispersed body of non-controlling shareholders in large corporations, the agency relationship practically evolved into a dual agency arrangement of shareholders-to-Board of Directors (BOD), and BOD-to-the-CEO. Given the inherent separation of ownership and control in this formulation, it is presumed that independent BODs will act in the shareholders' best interests and minimize opportunism on the part of management, and that they will monitor and restrain a CEO's managerial powers with an incentive and reward system related to a specific level of performance (Main et al. 1995).

Under the agency theory, the pay–performance rewarding policy will best align the CEO and the shareholders' interests (Jensen & Meckling 1976, Fama & Jensen 1983). Performance-based pay structures were classified as *market-based performance measures,* such as earnings (Murphy 1999), return on equity (Sigler 2011), return on assets (Ledford and Lawler, 2018), share price performance (Kirstin 2018), measurement of a CEO's skills (Daines 2005), market-to-book, return on assets, and shareholders' returns (Devers et al 2007), and *accounting-based performance measures,* such as debt structure, inventory management and accounting procedures (Devers et al. 2007), and cash flow (Nwaeze et al. 2006). Examining evidence from these descriptive studies into CEO compensation reveals insignificant, sometimes negative, correlations to corporate performance, and further suggests that incentives are mostly independent of stated monitoring metrics. Main et al (1995) have noted that social and psychological influences on the BOD, rather than economic metrics, may be responsible for significant increases in CEO compensation. Bolton et al (2006) and Core et al (1999) suggested that weak BODs contribute to rent extraction by the CEO; Malmendier and Tate (2007) noted rent extraction by "award winning" CEOs. Aguinis et al (2018) uncovered a clear decoupling between CEO pay and performance, as overperforming CEOs are underpaid and underperforming CEOs are overpaid. CEOs are often paid for luck and bad performance (Daines et al. 2005, Bertrand and Mullainath 2000) and charisma (Tosi et al. 2004). A recent report in *The Wall Street Journal* found that for the fourth year straight, the biggest US companies set CEO pay records even as a majority delivered negative stock-market returns to their shareholders (Francis and Fuhrmans 2019). The only meaningful correlation was observed between CEO and BOD pay, as noted by Brick et al. (2006).

What lead to this seemingly miscarried compensation policy? Researchers submitted contributing factors such as weak BODs, the "capture" of BODs by the CEO, the cozy relationships between CEOs and the board (Brick et al.

2006) and CEO duality (acting as CEO and Chairman of the Board)—just to name a few. Information asymmetry may entice unethical CEOs to cheat BODs by manipulating operational results for personal gain, or by reducing the probability of being replaced (Harris and Bromiley 2007, Schweitzer at al. 2004, Sigler 2011). Another contributing factor is tied to institutional investors' inclination to shy away from BOD participation, notwithstanding their 70% ownership of the stock market. Tirole (2001) noted the aversion of US financial institutions towards directorship appointments because of regulatory and fiscal rules (such as those on diversification), and because interpretations of the insider trading regulations penalize the resell of shares by monitoring institutions.

The mechanics of CEO pay have grown ever more complex, and the policy rules remain simple; star performers get a raise, and so do most of the rest (Francis and Fuhrmans 2019). The very large sums paid to CEOs have drawn the attention of economists (Bebchuk and Fried 2003), accountants (Bettis et al 2018, Lambert and Larker 1987), organizational behaviorists (Tosi et al 2000, Zajac at al. 1994), management policy (Baker et al. 2019, Devers et al. 2007, Zhang et al. 2008), governance policy (Core et al, 1999, Edmans et al. 2019, Papadopoulos 2019) and legal regulators (Baker et al. 1988, Booth 2005, Chang et al. 2010, Rose and Wolfram 2000). Most studies noted decoupling between CEO total pay (salary, bonus, stock, options and other pay) and corporate performance. The BOD appoints an executives compensation committee to compute the CEO pay. The compensation committee relinquishes this responsibility to a compensation consultant, thereby creating a new layer of the agency relationship. The role of compensation consultants and their widely-used peer group benchmarks is covered by the excellent work of Bizjak et al. (2009), who criticized benchmarking because it provides a mechanism to reward CEOs that is independent from firm performance. The author's survey of compensation consultants employed in the sample companies found an exclusive group of only a few consulting firms that are employed by most companies. This concentration, and the fact that many directors serve on multiple BODs, probably plays a role in both compensation convergence and the acceleration in average CEO compensations. This is analogous to all NBA players being compensated like LeBron James.

Figure 1 depicts the growth of the average CEO's compensation trend and the reduced variability over time in average compensation, as measured by the standard deviation bands above and below the average CEO compensation. The average compensation linear regression line represents the average of 100 CEO compensations for each year. We also calculated the standard deviation of the same group for each year and plotted linear regression bands (upper and lower lines) representing one standard deviations above and below the mean. If the reduction trend in standard deviation values across the sample data set continues, variability is expected to diminish in twenty years, when all CEOs will be paid a singular compensation – the average.

Additionally, the studies noted above share a common fundamental timing hypothesis that states CEO compensation is tied (or not) to a firm's performance through the same 12 month time frame during which performance is measured. This assumption is inherently faulty, as it employs annual accounting metrics that do not consider outcomes of long-term strategic actions and leadership qualities such as vision, decisiveness, and competence as examples. The 12-month performance-compensation cycle puts CEOs under constant pressure for short-term deliverables ranging from days (stock price movement) to quarterly results. Additionally, the push for short-termism negatively affects capital investment, R&D efforts and margins, and can lead to earnings management and stock price manipulations with long-term consequences beyond the current CEO's tenure.

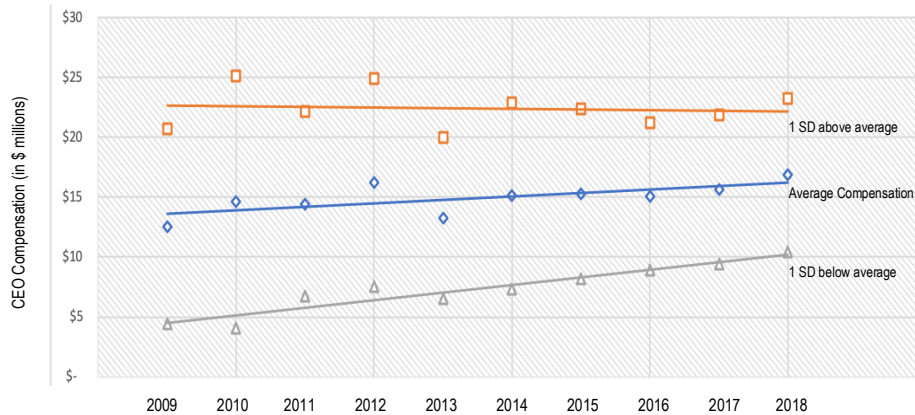

*Figure 1: Average CEO Compensation 2009-2018*

It is clear why most of the noted studies above failed to find a correlation between CEO compensation and performance: there isn't one. Compensation construct is independent of performance and mostly influenced by compensation consultants' benchmarks and BOD-CEO relationships. Barton et al. (2017) suggested a descriptive analysis based on a certain set of metrics to prove that long-term oriented companies deliver superior future results. This view was adopted by some of the major institutional investors like Blackrock and Vanguard, who found the concept a better match for their long-term investment horizon. They were quick to embrace the long-termism concept and suggested that companies in which they hold a stake should promote long-termism (Fink 2019 and Vanguard 2019).

The Barton study was conducted on 650 publicly-traded US corporations, with or without controlling shareholders. We posit that it is important to differentiate between controlled and non-controlled firms. Separation of ownership and control in controlled firms does not fit the definition of a pure agency relationship; large blockholders can control the board of directors and governance may be less independent. Furthermore, blockholders may also worsen governance by extracting private benefits from this control, or by pursuing objectives other than firm value maximization (Edmans 2014). CEO selection, compensation and performance metrics design are greatly influenced by the controlling entity. As a result, compensation of CEOs may be totally independent of performance and may vary from $1 to hundreds of millions per year (Loureiro 2014, Clifford 2017), risk-taking behavior can significantly affect a firm's performance (Wright et al. 1996), and major corporate decisions are decided differently in the presence of large blockholders (Holthausen 2003). For these reasons, we decided to limit our study to non-controlled companies.

We found Barton's concept of long-termism interesting as it suggests a link between measurable current metrics and long-term corporate performance. If this concept holds true in non-controlled long-term corporations, and if there is a link between future CEO compensation and long-term corporate management goals (see Barton's et al. model), it may provide a novel approach to setting CEO compensation in order to optimistically affect future corporate performance.

**2. Data**

The study assembled a database by extracting financial information regarding publicly-traded companies on the S&P 500 index with continuous operations from 2009 to 2018. Data sources include SEC Form DEF 14A (also referred to as the Proxy Statement) and SEC 10K (Annual Report) that are filed annually with the Securities and Exchange Commission (SEC). Our data is limited to companies without any blockholders (defined as stock ownership >10%). The list of such companies was provided by Kazaz (2020), and excluded companies with a single class of shareholders to prevent potential bias due to controlling shareholders. We also excluded companies that either went through a merger, or new companies that have been in existence for less than 10 years in order to remove longitudinal data gaps.

The Barton et al. model of long-termism suggested a set of independent variables to predict the dependent variables measuring future performance. We employed a similar set of variables, with the exception of one independent variable measuring external input based on analysts' earnings target. We also added CEO compensation to the list of independent variables to test the hypothesis that it may predict future corporate performance. Finally, we added total shareholder return to the response variables in order to test its predictability by the explanatory variables.

Table 1: Explanatory Variables

| variable | abbr. | definition |
|---|---|---|
| Investment Ratio | IR | Ratio of capital expenditures to depreciation |
| Earning Quality Ratio | EQ | Accruals as share of the revenues |
| Margin Growth Ratio | MG | Earning growth divided by revenue growth |
| Earnings per ShareEPS | EPS | EPS Growth less True Earnings Growth |
| CEO Total Package | CEOt | Salary, bonus, stock, stock options and other pay |

Table 2: Response Variables

| variable | abbr. | definition |
|---|---|---|
| Revenues | REV | Gross sales less discounts |
| Earnings | Earn | Earnings available to shareholders |
| Economic Profits | Eprof | Net Income less Opportunity Costs |
| Market Capitalization | MCap | Year-end Share Price times # of diluted shares |
| Total Shareholder Return | TSR | Capital gain plus dividend per share |

## 3. Methodology and Results of the Data Analysis

### 3.1 Methodology

The analysis here uses Canonical Regression Quantiles to produce an index based on past data in order to predict the response variables in the future (see Portnoy and Haimberg 2020). A classic approach would be to use Canonical Correlations, which find the linear combination of explanatory variables that is most highly correlated with a linear combination of the responses. Unfortunately, Canonical Correlation analysis is developed under the assumption that all variables follow a multivariate, normal distribution. Inference for Canonical Correlations can be very poor if the data is skewed, heavy-tailed, contains outliers, or suffers from other departures from normality. The Canonical Regression Quantile approach ameliorates these shortcomings substantially. Furthermore, by focusing on quantiles rather than means, it allows for an analysis that disaggregates the responses according to their size.

The method of Canonical Regression Quantiles seeks to find a linear combination of the response variables that is best predicted by a linear combination of the explanatory variables in terms of regression quantile objective functions. The basic idea is as follows: let $\rho_\tau(u) = u(\tau - I(u < 0))$ be the quantile objective function so that the $\tau$th quantile, $Q_\tau$ of a sample $\{x_i\}$ satisfies

$$Q_\tau = \min_\theta \sum_i \rho_\tau(x_i - \theta) \tag{1}$$

Given a data matrix, $X$, of explanatory variables (generally including a constant intercept column) and a data matrix $Y$ of response variables, we would like to define the first Canonical Regression Quantile as the pair of coefficient vectors, $(\alpha, \beta)$ achieving

$$\min_{\alpha,\beta} \sum_i \rho_\tau(x_i'\beta - y_i'\alpha) \tag{2}$$

where $x_i$ and $y_i$ denote the $i$th rows of $X$ and $Y$ respectively. Unfortunately, this will not define unique solutions since the objective function in (2) is invariant under scale changes. In the case of Canonical Correlations, this problem is resolved by imposing orthogonality (or, more generally, a quadratic norm identity). However, to reflect the quantile setting and to avoid the lack of robustness of quadratic measures, it would be preferable to impose the following identity:

$$\sum_j |\alpha_j| = 1 \tag{3}$$

where $\alpha_j$ are the coordinates of $\alpha$. If all the $\alpha_j$ are positive, (3) is equivalent to $\sum_j \alpha_j = 1$, and minimizing (2) subject to this restriction becomes an easily-solvable linear programming problem.

A more complete development of this approach, together with a derivation of basic inferential properties, are given in Portnoy and Haimberg (2020). At this point, we refer the reader to this paper, and carry out the analysis in terms of the linear combination of explanatory *X*-variables (which we call an index) that best predicts a linear combination of the response *Y*-variables in terms of the regression quantile objective function. The analyses here will address three questions: (1) can an index based on past data be constructed so that it predicts future economic performance better than current CEO total compensation (CEOtot), (2) can this analysis suggest under- and over-compensation for CEO's, and (3) does the excess of CEOtot over its prediction using the index bear a relationship with economic performance?. That is, does the residual in predicting current CEOtot from the index correlate with economic performance? Note that positive residuals mean CEOtot exceeds what past performance would predict, and smaller residuals mean that CEOtot was set lower than what past performance would indicate. Thus, if the residual relates positively to future economic performance, then the method of setting CEOtot is adding to economic performance. Otherwise, the setting of CEOtot is not (positively) affecting performance. If the residuals relate negatively to economic performance, then better performance could be obtained by basing CEOtot on the index.

### 3.2 Construction of an index

For the data described above, a "long-term" perspective suggests using several years of data to predict a small number of years ahead. Given only 10 years of data, we focus on using five years' worth of data to predict two years ahead. Past values of both *X* and *Y* variables will be used, perhaps suggestive of "Granger Causality" (Granger, 1980). In addition, we will begin by including past CEO total compensation. Assuming we have observed seven years, we use the Canonical Regression Quantile method above and classical Canonical Correlations to find linear predictors (indices given by $\beta$-coefficients) based on the past 5 years of explanatory variables to predict the Y-variables in the current year.. That is as described in detail below, we use the $\beta$-coefficients for the first five years' data to produce a prediction index, then use this index to predict two years into the future. The indices produced will then be compared with current CEOtot in terms of predicting future *Y*-variables.

With only 100 companies, it is not reasonable to use all the *X* and *Y* data over five years. From the long-term perspective, we may consider replacing each variable by simple aggregates of the five years of data. Specifically, we consider two such aggregates: a discounted average, discounting 5% each year into the past (essentially an exponential smoothing); and the minimum of the four differences between successive years. The first measures overall performance while the second measures some form of stability. This leads to two measurements for each of the four *X* variables, the five response Y variables, and CEOtot: a total of 20 variables. In addition, we introduce an indicator variable for each of the six industry types (implicitly including an intercept). This creates a 100-by-26 matrix of explanatory variables (for each time period) that we denote using $X^*$. Letting $Y^*$ denote the response variables two years ahead of the data for $X^*$ (that is, two years ahead of the "current" year), we apply two "Canonical" analyses to develop predictor (*X*) and response (*Y*) indices. The first is the Canonical Regression Quantile method described above, and the second uses the vectors corresponding to the leading (classical) Canonical Correlation. We then use these indices applied to the data from the current year and the previous four years to predict each of the five *Y*-values two years ahead. To be consistent with the basic formulations, we use Least Squares to predict two years ahead for classical Canonical Correlation index, and we use median ($L_1$) regression to predict for the Canonical Regression Quantile index here. We then compare the predictions with the observed *Y*-values (two years ahead).

To be precise, for example, we use the *X* and *Y* data from years 2009 to 2013 to predict *Y* data in 2015 as a means of creating the index coefficients, and then apply these coefficients to data from years 2011 to 2015 to predict *Y* variables in year 2017 (by median regression, that is, using a regression quantile estimator with proportion $\tau = .5$). We then repeat this process for data from 2010 to 2014 to create an index for predicting *Y* variables in year 2018. Of course, other spans of time can be used, both for creating the predictor indices and for predicting into the future. As remarked after presenting the statistical results below, runs using other choices were done. Though the details of the alternative analyses differed, there were no substantive differences in conclusions.

As a final remark concerning the data set-up, note that like typical economic measurements, the *Y* variables were expected to be generally positive values but with clear non-normal variability. Thus, we followed common economic practice and took logs. Unfortunately, some values were actually negative, especially for the variables "Economic Profits" and "Total Shareholder Returns." After perusing the data, we decided to use a signed log-transformation: sign(y) log |y|, which seemed to be indicated by the rather large variability in the negative values for Economic Profits. Some runs with the more usual log(max(1, *y*)) transformation were done, and they showed almost no difference (generally in the third significant figure or less).

To analyze the results, we consider three predictors: the index predictors using each of the two Canonical methods and also predictions based on CEOtot. To allow consistent comparisons, we use both Least Squares and Median regression to predict performance (two years ahead) based on CEOtot. To elucidate the Canonical Regression Quantile method, we will look at the *α* and *β* coefficients and their standard errors based on the weighted version of Andrews' bootstrap described in Portnoy and Haimberg (2020). Numerical comparisons were made in terms of root-mean-squared-error (RMSE) and the mean-absolute-error (MAE), combining results for years 2017 and 2018. However, only the MAE results are reported here, as MAE is more robust and less sensitive to non-normality. We assess statistical variation in these values using the same weighted version of the Andrews' bootstrap. We also look at some graphical presentations to get a more intuitive idea of the issues, focusing on the Canonical Regression Quantile and the CEOtot predictors. Among the plots considered, we present here scatter plots of predictor index (from the Canonical Regression Quantile approach) and CEOtot, versus the Canonical Regression Quantile response index two years ahead. We also present scatter plots of the index and CEOtot versus each of the response variables separately. Finally, we present plots of *y* vs. *ŷ* in terms of Q-Q plots. All computer work was done in R (R core team, 2015). Median regression used the function `rq` from the R-package, `quantreg`, while Least Squares regression used the base function `lm`. The results are as follows. The *α* coefficients for the "best" predicted combination of response variables are given in Table 3.

Table 3: α coefficients

| Year | logRev | logEarn | logEprof | logMarCap | logTSR |
|------|--------|---------|----------|-----------|--------|
| 2017 | .884   | .003    | 0        | .113      | 0      |
| 2018 | .998   | 0       | 0        | 0         | .002   |

Clearly, $\alpha$ is controlled almost entirely by log Revenue, and it would likely be adequate to base $\beta$ on the regression of log Revenue alone on $X^*$. This was done as a check on the Canonical Regression Quantile approach, since standard bootstrap inference for conditional quantile analysis is legitimate. As expected, the differences between using log Revenue alone and the Canonical Regression Quantile index were very small, generally in the third significant figure or smaller. The $\beta$ coefficients and their t-statistics (using the weighted bootstrap standard errors) are given in Table 4.

Table 4: Beta coefficients for Index

|            | 2017    | t-stat  | 2018    | t-stat  |
|------------|---------|---------|---------|---------|
| intercept  | 0.754   | 1.529   | 0.022   | 0.046   |
| indust     | 0.113   | 0.886   | 0.217   | 2.110   |
| health     | 0.280   | 1.888   | 0.320   | 2.853   |
| consum     | 0.097   | 0.651   | 0.129   | 1.103   |
| energy     | 0.277   | 1.402   | 0.402   | 2.086   |
| tech       | 0.264   | 1.678   | 0.3212  | .240    |
| IRwt       | -0.026  | -0.830  | -0.020  | -0.739  |
| IRminD     | -0.013  | -0.236  | -0.024  | -0.421  |
| EQwt       | -0.003  | -0.755  | 0.002   | 0.408   |
| EQminD     | -0.001  | -0.266  | -0.008  | -1.499  |
| MGwt       | -0.006  | -1.255  | -0.006  | 1.341   |
| MGminD     | -0.0005 | -0.179  | 0.0002  | 0.096   |
| EPSGwt     | -0.001  | -0.308  | -0.001  | -0.248  |
| EPSGminD   | 0.0002  | 0.187   | -0.0001 | -0.051  |
| CEOtwt     | 0.001   | 0.156   | -0.011  | -1.427  |
| CEOtminD   | -0.0001 | -0.084  | -0.004  | -1.006  |
| logRevwt   | 0.896   | 7.028   | 1.059   | 11.821  |
| logRevminD | 0.859   | 3.516   | 0.994   | 4.553   |
| logEarnwt  | -0.052  | -0.738  | -0.064  | -1.088  |
| logEarnminD| 0.017   | 0.935   | 0.020   | 1.281   |
| logEprofwt | 0.004   | 0.438   | 0.001   | 0.101   |
| logEprofminD | -0.016 | -1.946 | -0.008  | -1.157  |
| logMCapwt  | 0.048   | 0.386   | -0.010  | -0.097  |
| logMCapminD| -0.180  | -0.671  | -0.178  | -0.720  |
| logTSRwt   | 0.032   | 2.079   | 0.012   | 0.737   |
| logTSRminD | -0.004  | -0.628  | 0.001   | 0.114   |

It is somewhat surprising that many of these coefficients are significantly negative, but also gratifying that many are significantly non-zero (or nearly so). Most also seemed somewhat consistent from 2017 to 2018. These results suggest strongly that there is a signal. This is confirmed by Table 5, which gives values of the mean absolute error (MAE) averaged over the two years of prediction.

Table 5: MAE: 2017-8 mean (SD)

| Method | logRev | logEarn | logEprof | logMCap | logTSR |
|---|---|---|---|---|---|
| cancor | 0.190 (.049) | 1.878 (.271) | 6.050 (.209) | 0.498 (.042) | 7.036 (.192) |
| rq.can | 0.162 (.028) | 1.413 (.020) | 4.940 (.065) | 0.529 (.027) | 6.262 (.162) |
| CEOrq | 0.776 (.012) | 1.610 (.015) | 5.081 (.142) | 0.658 (.014) | 6.327 (.170) |

Clearly, the Canonical Regression Quantile index (rq.can in the table) provides better predictions than CEOtot for all response variables under the more robust MAE criterion. The plots below show that the moderately large number of negative values for log Economic Profits and Total Shareholder Return make these variables somewhat anomalous, and this point will be discussed further below. Finally, some plots for the 2017 predicted responses may clarify the nature of the data. Plots for 2018 are quite similar. Figures 1 and 2 present scatter plots of the Canonical Regression Quantile index and CEOtot on the x-axis against the Canonical Regression Quantile $Y$-index ($\alpha'Y$, see Figure 2) and against each of the five $Y$ variables (Figure 3). The median regression lines are also plotted.

It is clear that the Canonical Regression Quantile index is predicting its $Y$-index much better than CEOtot. Considering the plot for logEprof and logTSR, clearly the negative values are well-separated from positive ones, and are remarkably large in absolute value and rather variable. Both indices predict the more numerous positive values quite well (as would be expected for a median predictor), but the CEOtot predictions seem pulled down more toward the negative values, and thus tend to underestimate the positive ones. There are only a few negative responses for other variables, and so these act more like outliers.

Figure 3 gives Q-Q plots of the observed $Y$-response vs. the predicted response from each of the two methods. The separation of negative and positive responses is clear again, as is the tendency of CEOtot to overestimate smaller responses and underestimate larger ones.

To provide a numerical measure of how much better the index does as compared to CEOtot in predicting the $Y$-variables, we consider predicting each $Y$ variable from both CEOtot and the index together (in terms of median regression). For example, consider using the index (defined from $X$-data in 2009-2013 and current $Y$-outcomes in 2015, (and 2015 CEOtot), to predict log(Earnings) in 2017. The results for median regression are summarized in Table 6.

Table 6: Log(Earnings) vs. CEOtot and Index

| Coef | Value | Std. Error | t-value | Pr( > |t|) |
|---|---|---|---|---|
| CEOtot | .00363 | .00272 | 1.335 | .091 |
| CEOrq | .9093 | .0257 | 35.38 | .000 |

Note that the *t*-statistic for the index is much larger that the *t*-statistic for CEOtot, strongly indicating that the index is a much better predictor. To summarize all such regressions (for all $Y$-variables and all possible years), we computed the ratio of the *t*-statistic for the index to the t-statistic for CEOtot. In fact, nearly half the CEOtot coefficients were negative. The absolute values of the ratio varied from 1.63 to 235 with a median of 4.85 and first and third quartiles equal to 2.57 and 21.5. This is extremely strong evidence that the Index predicts future economic performance much better than CEOtot.

**Remarks:** A number of particular specifications were employed in the analysis summarized above. In fact, many of these specifications were generalized in trials not reported here. Several runs considered using four years of prior data to predict one, two, and three years ahead. The results differed in detail, but were qualitatively the same.

Notice that the standard errors for the for logEprof and logTSR are somewhat larger than other standard errors, reflecting the bimodality of this data. Nonetheless, the standard error estimates above seem reliable. Using the standard weighted bootstrap gave similar results, as did the standard bootstrap using total revenue alone to develop the index. Using a larger resampling size ($R = 600$ rather than $R = 200$) also gave similar standard error estimates, with the same cases of larger standard errors.

The original idea was to use only the *X*-data to develop the indices. Some runs were carried out using only the *X*-data (all the *X*-data and not just the two summary aggregates) from two to four earlier years. This approach did not produce very good predictors. Some attempts at variable selection were also tried. In almost all cases, the variable selection methods suggested that all variables were needed, and so differed very little (or not at all) from the results presented here.

### 3.3. Do current methods of setting CEOtot improve future economic performance?

Having found an index that predicts future economic performance better than CEOtot, can this index be used to improve the process of setting CEO compensation? To begin, we ask how well the index predicts CEOtot. For this purpose (and in the remainder of this section) the index will be defined as described in the previous section, except that past values of CEOtot will not be used in the *X*-variables used to define the index.

We now consider how to use the predictability of CEOtot in order to improve its determination. Again, we focus on the index that uses five consecutive years of *X*-data (as described above) to predict the "optimal" combination of *Y*-variables two years ahead. Specifically, we consider whether the CEO was overpaid or underpaid with respect to the prediction of current CEOtot from the index. In particular, we look at the difference between the current value of CEOtot and its prediction from the index (which is based on the five-year span ending two years ago). This difference is just the residual, CEOres, from fitting CEOtot from the index using a median regression fit (that is, a regression quantile estimate with $\tau = .5$ ). Again, we justify median regression by noting the lack of normality of the data and the robustness of regression quantile methods.

A positive value of CEOres indicates that CEO compensation was set higher than what would be suggested by the index, and negative values indicate the opposite. If CEOres is (significantly) related to future economic performance, the method used to set CEO compensation is clearly adding to future performance; while is it is negative, it would clearly be better to base CEO compensation on the index directly. If the relationship between CEOres and future performance is not significantly different from zero, then the present methods of determining compensation are not helping future performance, but the index itself cannot add any improvement. Here are the results of predicting future performance from CEOres. Table 7 lists the *t*-statistics for predicting each of the *Y*-variables from CEOres using median regression. The standard error used to define the *t*-statistic is based on the standard bootstrap (with 600 replications) in the `rq` function in the `quantreg` package. This can be justified here since (at least asymptotically) the residuals are independent of the regression estimates and thus the median regression of a *Y* variable on CEOres should be approximately independent of the index.

The *t*-statistics in Table 7 are generally moderately small. To assess significance, we must adjust for multiple tests. There are 25 *t*-statistics in the table, and so by the Bonferroni method, the critical value providing simultaneous significance at level .05 is (conservatively) the normal quantile at 1-.025/25, which is 3.09. While there is a suggestion that CEOres has some value in predicting logMCap, none of the *t*-statistics exceeds 3.09. Thus we can conclude that our data provides no evidence that the methods of determining CEO compensation gave improved economic performance two years ahead.

Table 7: *t*-stats for predicting outcomes from rq (CEO ~ Index, .5) $resid

| Index-yr | pred-yr | logRev | logEarn | logEprof | logMCap | logTSR |
|---|---|---|---|---|---|---|
| 2015 | 2016 | 0.466 | 1.840 | 0.114 | 1.290 | 0.687 |
| 2015 | 2017 | 0.838 | 0.772 | -0.446 | 1.680 | 0.458 |
| 2015 | 2018 | 1.110 | 1.690 | -0.214 | 2.250 | 1.080 |
| 2016 | 2017 | 0.823 | 1.990 | -0.214 | 1.940 | 1.070 |
| 2016 | 2018 | 0.939 | 0.021 | -1.340 | 2.400 | 0.464 |

**4. Discussion**

The use of an Index based on past data constructed with the Canonical Regression Quantile Method first suggested by Portnoy and Haimberg (2020) provides an exciting new window into long-termism and CEO compensation settings. Clearly the Index performs well in predicting future performance, but not in a significant way for predicting future CEO compensation (see figures 2-3). Our results suggest that the Index provides strong evidence that long-termism as measured by the input variables initially suggested by Barton et al. (2017) apply to the non-controlled subset of public corporations. We also explored the incorporation of CEO compensation into the Index to reliably predict future performance, but found poor signal when compared to the more powerful Index. This repudiates the belief that higher paid CEOs perform better (Daines et al. 2005), as we found that performance is not sensitive to CEO pay.

Lastly, while conceding the poor regression of future CEO compensation, we applied a previous five-year Index to predict next year's CEO compensation (for example using the data from 2010 - 2016 to construct Index16, and using Index16 to predict CEOtot17 and CEOtot18 from Index17). We also compared actual compensation vs. the predicted compensation. The resulting residual between actual compensation and the predicted compensation, expressed in percentages, markedly differentiates between over-and under-paid CEOs, as figures 5 and 6 demonstrate. This is a positive finding worthy of further research with a larger sample size than the 100 companies we used, and with a longer time series than our ten years. We suggest that future research with a larger data set could more finely-tune the Index, increase its sensitivity, and most likely result in a better prediction formula for CEO compensation.

## 5. Conclusion

The journey through previous literature on CEO compensation revealed no suggestive correlation between CEO compensation and firm performance. Additionally, a trend analysis of CEO compensation (see Figure 1) demonstrated a noticeable convergence in the variability of CEO compensation across our sample of 100 companies over the past ten years, possibly due to compensation committees' reliance on shared compensation consultants and peer group comparisons (Bizjak 2009). This convergence, together with other non-economic factors such as BOD-CEO relationships, is clearly suggestive of the influence factors other than performance have on compensation . As no direct association was found the in previous literature, we seek to establish an indirect approach to identify companies with superior future performances, and to associate CEO compensation with identifiable present input variables found to predict superior future performance.

Barton et al. (2017) found that companies who practice "long-termism," as measured by a group of independent variables, post long-term superior financial and market results. We propose a novel approach using the Canonical Regression Quantile Method to answer three questions: Is long-termism applicable to non-controlled public companies? Can current CEO compensation influence future performance? And finally, is it possible to predict future CEO pay from the long-termism independent variables?

An Index based on past long-termism data was constructed with the Canonical Regression Quantile Method, as first suggested by Portnoy and Haimberg (2020), provides an exciting new window for answering those questions. The Index focuses on predicting two years into the future based on five years of past data (which accords with a long-term view). The results confirm the long-termism concept and reinforce Barton's findings in non-controlled public corporations. The inclusion of past CEO compensation in the Index produced a weak signal, and we found no evidence that current CEO compensation may influence future performance. Put simply: overpaying the CEO will not by itself produce superior financial results.

The Index's prediction of future CEO compensation demonstrated its ability to determine if CEOs are over- or under-paid with a low level of confidence. While the results are preliminary, they strongly support the potential value of the Canonical Regression Quantile approach. Further research of the use of this approach on a larger data set may increase the Index sensitivity and produce a formula to calculate fair compensation based on long-termism considerations, free from the uncertainty compensation committees that introduce into the CEO compensation-setting process.

.

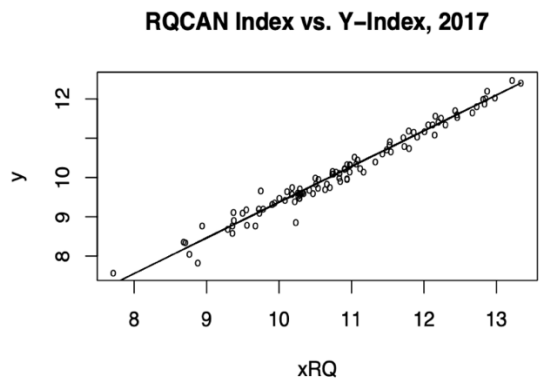

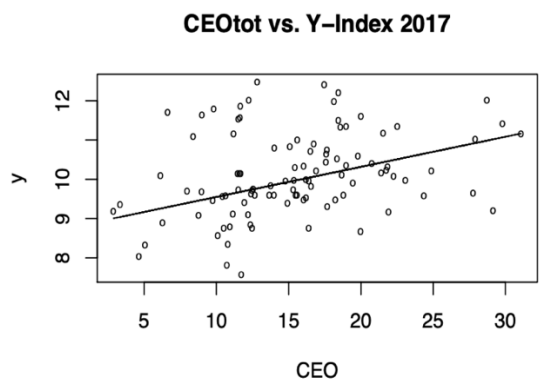

Figure 2: 2017: response index vs. prediction index and CEOtot

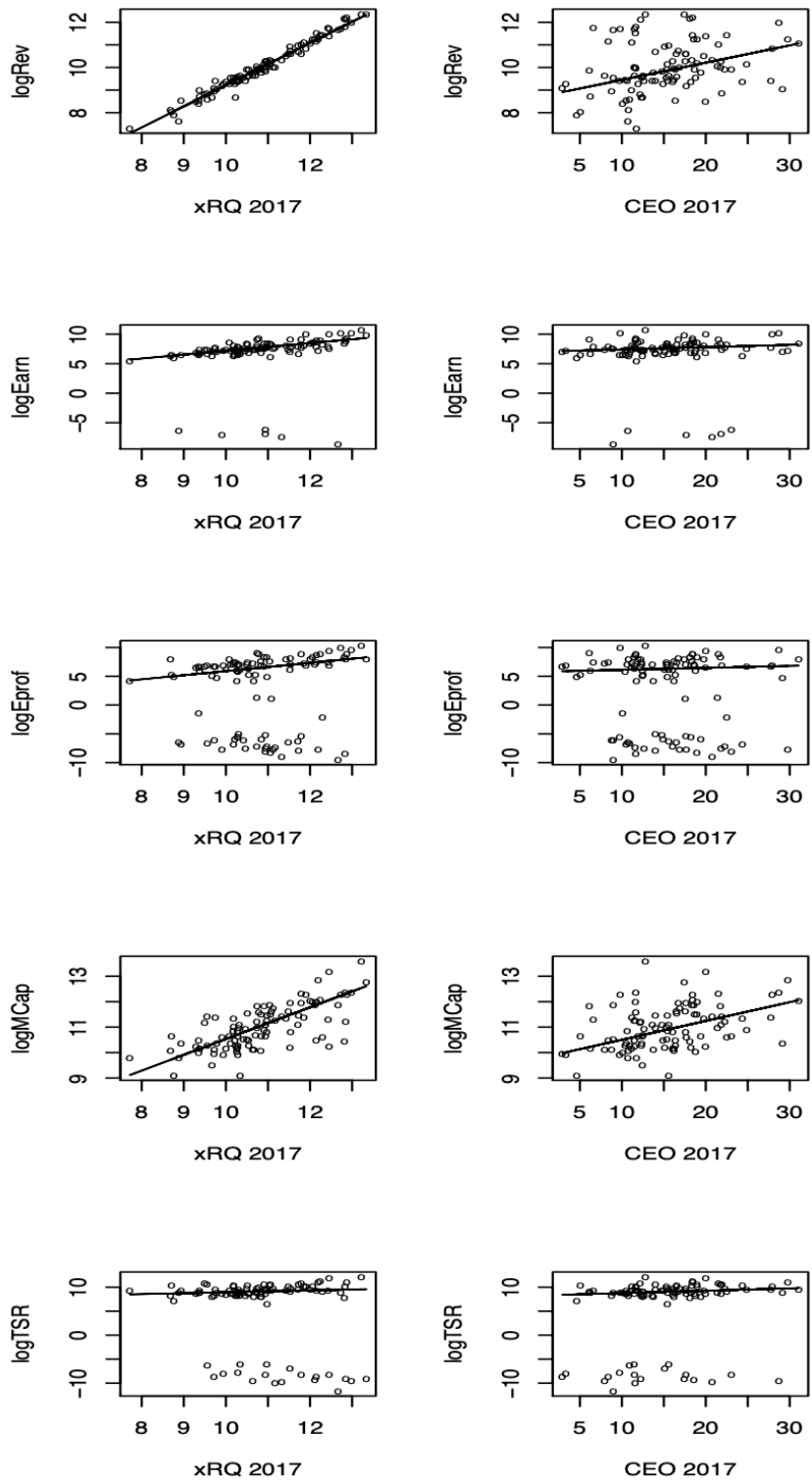

Figure 3: 2017: Y-variables vs. prediction index and CEOtot

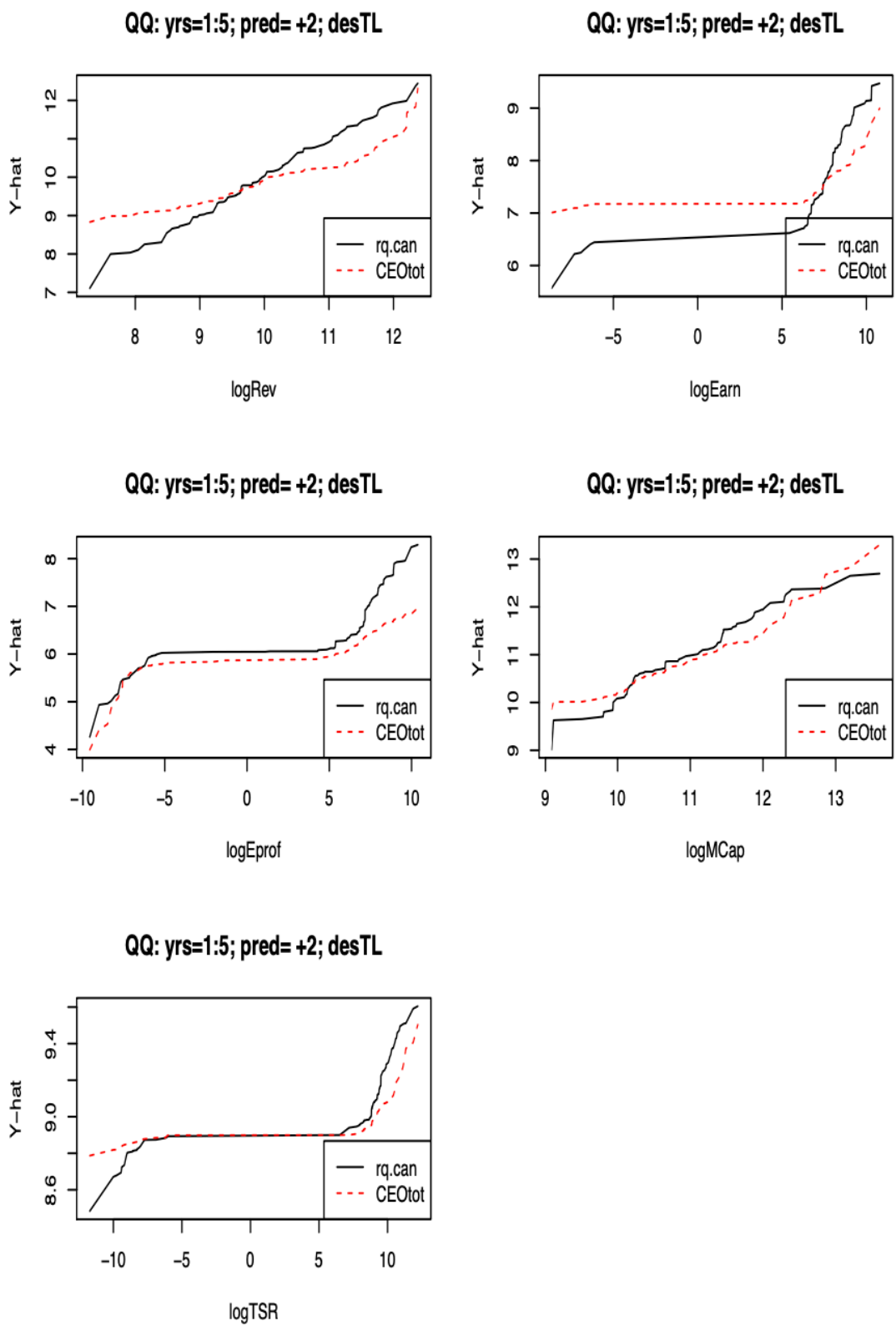

Figure 4: 2017: Q-Q plot, fit vs. response

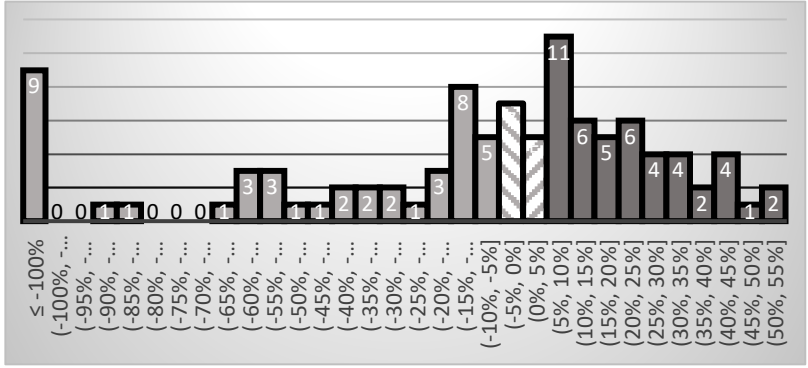

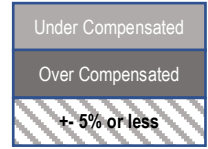

Figure 5: 2017 - CEO Distribution of Over and Underpaid Compensation

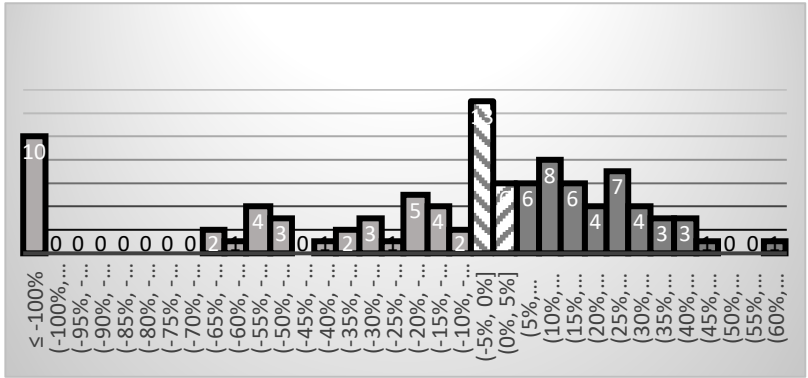

Figure 5: 2018 - CEO Distribution of Over and Underpaid Compensation